\documentclass[a4paper]{article}
\usepackage{amsmath}
\usepackage{amssymb}

\usepackage{amsmath,amssymb,amsfonts}
\usepackage{graphicx}

\newtheorem{theorem}{Theorem}[section]
\newtheorem{lemma}{Lemma}[section]

\def\Z{{\mathchoice{{\bf Z}}{{\bf Z}}{{\rm Z}}{{\rm Z}}}}

\def\R{{\mathchoice{{\bf R}}{{\bf R}}{{\rm R}}{{\rm R}}}}
\def\C{{\mathchoice{{\bf C}}{{\bf C}}{{\rm C}}{{\rm C}}}}
\def\S{{\mathchoice{{\bf S}}{{\bf S}}{{\rm S}}{{\rm S}}}}

\let\epsilon=\varepsilon
\let\phi=\varphi
\title{One-dimensional quantum walks with a general perturbation
of the radius 1}

\author{M.~Ryazanov\thanks{Mechanics and Mathematics Faculty, Lomonosov Moscow State University, Leninskie
Gory~1, Moscow, 119991, Russia, E-mail: mikhail.ryazanov@mail.ru} 
\and
A.~Zamyatin\thanks{Mechanics and Mathematics Faculty, Lomonosov Moscow State University, Leninskie
Gory~1, Moscow, 119991, Russia, E-mail: andrei.zamyatin@mail.ru}
}
\date{}

\begin{document}
\maketitle

\begin{abstract}
We give a complete description of the discrete spectrum of one-dimensional
Hamiltonian with a general perturbation of the radius $1$. 
\end{abstract}

\allowdisplaybreaks

\section{Introduction}

We consider the one particle continuous time one-dimensional quantum
walk with an Hamiltonian $H$ acting in the Hilbert space $l_{2}(Z),$
where $\Z$ is the one-dimensional lattice and $l_{2}(\Z)$ is the
space of complex valued square summable sequences $f=\{f_{n}\in\C,n\in\Z\}$.

Let $\left\{ e_{n}\right\} _{n\in Z}$ be the standard basis of $l_{2}(\Z)$.
If $f\in l{}_{2}(\Z)$ then we have $f=\sum_{n\in Z}f_{n}e_{n}$.

The Hamiltonian $H=H_{0}+H_{1}$, $l_{2}(\Z)\rightarrow l_{2}(\Z)$
is defined as follows:
\begin{align}
H_{0}e_{n} & =-\lambda\left(e_{n+1}-2e_{n}+e_{n-1}\right),\:n\in\Z\nonumber \\
H_{1}e_{0} & =-\lambda_{1}\left(e_{1}-2e_{0}+e_{-1}\right)+\mu e_{0}\nonumber \\
H_{1}e_{1} & =-\lambda_{1}e_{0}+\mu_{1}e_{1}\nonumber \\
H_{1}e_{-1} & =-\lambda_{1}e_{0}+\mu_{1}e_{-1}\nonumber \\
H_{1}e_{n} & =0,\text{if}\:|n|>1,\label{H5}
\end{align}
where $\lambda\text{, }\lambda_{1}\text{, }\mu,\,\mu_{1}\in\R$. Without
loss of generality we can assume that $\lambda>0.$ Our aim is to
describe the discrete spectrum of the Hamiltonian $H$.

The continuous time quantum walk evolution is defined as 
\[
f(t)=e^{-itH}f(0)\in l_{2}(\Z)
\]
where $f(0)$ is the initial state.

We generalize the result of \cite{Lak}, where authors investigated
the discrete spectrum of the Hamiltonian which is defined by two parameters
$\mu$ and $\mu_{1}$, $H=H_{0}+H_{1}$ $\,:\,l_{2}^{s}(\Z)\rightarrow l_{2}^{s}(\Z)$:
\begin{align*}
H_{0}e_{n} & =-\frac{1}{2}e_{n+1}+e_{n}-\frac{1}{2}e_{n-1}\\
H_{1}e_{0} & =\mu e_{0}\\
H_{1}e_{1} & =\mu_{1}e_{1}\\
H_{1}e_{-1} & =\mu_{1}e_{-1}\\
H_{1}e_{n} & =0,\text{ if\quad}|n|>1
\end{align*}
where $l_{2}^{s}(\Z)\subset l_{2}(\Z)$ is the subspace of symmetric
sequences: $f_{n}=f_{-n}$ for all $n\in\Z$ .

\section{Main results}

\paragraph*{Notation}

It will be convenient to introduce new parameters 
\[
\alpha=\frac{\lambda_{1}}{\lambda}, \quad \delta=\frac{\mu_{1}}{\lambda},\quad \sigma=\frac{2\lambda_{1}+\mu}{2\lambda}
\]
Define the functions of parameters $\alpha,\delta,\sigma$:
\begin{align}
c_{1} & =c_{1}(\alpha,\delta,\sigma)=(\delta-1)(\sigma-1)-(\alpha+1)^{2}\\
c_{2} & =c_{2}(\alpha,\delta,\sigma)=(\delta+1)(\sigma+1)-(\alpha+1)^{2}
\end{align}
where $\left(\alpha,\delta,\sigma\right)\in\R^{3}$.

Let us introduce the following partition of $\R^{3}$:
\begin{align}
U_{1}(+,+)= & \left\{ \left(\alpha,\delta,\sigma\right)\in\R^{3}:c_{1}>0,c_{2}>0,\delta>1,\sigma>1\right\} \nonumber \\
U_{0}(+,+)= & \left\{ \left(\alpha,\delta,\sigma\right)\in\R^{3}:c_{1}\geqslant0,c_{2}\geqslant0,|\delta|\leq1,|\sigma|\leq1\right\} \nonumber \\
U_{-1}(+,+)= & \left\{ \left(\alpha,\delta,\sigma\right)\in\R^{3}:c_{1}>0,c_{2}>0,\delta<-1,\sigma<-1\right\} \nonumber \\
U(-,+)= & \left\{ \left(\alpha,\delta,\sigma\right)\in\R^{3}:c_{1}<0,c_{2}\geqslant0\right\} \cup\nonumber \\
 & \cup\left\{ \left(\alpha,\delta,\sigma\right)\in\R^{3}:c_{1}=0,c_{2}>0,\delta>1\text{,\ensuremath{\sigma}>1}\right\} \nonumber \\
U(-,-)= & \left\{ \left(\alpha,\delta,\sigma\right)\in\R^{3}:c_{1}<0,c_{2}<0\right\} \nonumber \\
U(+,-)= & \left\{ \left(\alpha,\delta,\sigma\right)\in\R^{3}:c_{1}\geqslant0,c_{2}<0\right\} \cup\nonumber \\
 & \cup\left\{ \left(\alpha,\delta,\sigma\right)\in\R^{3}:c_{1}>0,c_{2}=0,\delta<-1,\sigma<-1\right\} \label{U2}
\end{align}
Figures \ref{fig1}--\ref{fig3} (see Appendix) show sections of the subsets from partition
(\ref{U2}) by planes $\alpha=c$ for various constants $c\neq-1.$

Finally we put 
\begin{equation}
\kappa=\frac{\lambda(\delta+1)^{2}}{\delta}=\frac{(\mu_{1}+\lambda)^{2}}{\mu_{1}}\label{kappa}
\end{equation}
for $\mu_{1}\neq0.$

\paragraph*{Results}

It follows from Weyl's theorem \cite{Reed-4} on essential spectrum
that the essential spectrum of the Hamiltonian $H$ is the same as
for the homogeneous Hamiltonian $H_{0}$ and coincides with the segment
$[0,4\lambda].$

Denote by $s_{d}(H)$ the discrete spectrum \cite{Reed-1} of $H.$
We give a complete description of $s_{d}(H).$ We prove that the discrete
spectrum consists of at most three eigenvalues. One of them we find
explicitly. For the other two we define the intervals of the real
axis containing these eigenvalues.

\begin{theorem}\label{th1} Let $\lambda>0$ and $\kappa$ be defined
by (\ref{kappa}). For $\alpha=\frac{\lambda_{1}}{\lambda}\neq-1$
we have

$\text{a)}$ if $\left(\alpha,\delta,\sigma\right)\in U_{1}(+,+),$
then $s_{d}(H)=\{\nu_{1},\nu_{2},\nu_{3}\},$ where $\nu_{3}=\kappa,$
$\nu_{1},\nu_{2}\in(4\lambda,\infty)$ and, moreover, 
\[
  4\lambda<\nu_{1}<2\lambda\Bigl(\min\Bigl\{ \sigma,\frac{\delta^{2}+1}{2\delta}\Bigr\} +1\Bigr)\leq
  2\lambda\Bigl(\max\Bigl\{ \sigma,\frac{\delta^{2}+1}{2\delta}\Bigr\} +1\Bigr)<\nu_{2};
\]

b) if $\left(\alpha,\delta,\sigma\right)\in U_{0}(+,+),$ then $s_{d}(H)=\emptyset$;

c) if $\left(\alpha,\delta,\sigma\right)\in U_{-1}(+,+)$ then $s_{d}(H)=\{\nu_{1},\nu_{2},\nu_{3}\},$
where $\nu_{3}=\kappa,$$\nu_{1},\nu_{2}\in(-\infty,0)$ and, moreover,
\[
  \nu_{1}<2\lambda\Bigl(\min\Bigl\{ \sigma,\frac{\delta^{2}+1}{2\delta}\Bigr\} +1\Bigr)\leq
  2\lambda\Bigl(\max\Bigl\{ \sigma,\frac{\delta^{2}+1}{2\delta}\Bigr\} +1\Bigr)<\nu_{2}<0;
\]

d) if $\left(\alpha,\delta,\sigma\right)\in U(-,+),$ then

d1) for $|\delta|\leqslant1$ $s_{d}(H)=\{\nu_{1}\},$ where $\nu_{1}\in(4\lambda,\infty)$,

d2) for $|\delta|>1$ $s_{d}(H)=\{\nu_{1},\nu_{2}\},$ where $\nu_{1}\in(4\lambda,\infty),$$\nu_{2}=\kappa;$

e) if $\left(\alpha,\delta,\sigma\right)\in U(-,-),$ then

e1) for $|\delta|\leqslant1$ $s_{d}(H)=\{\nu_{1},\nu_{2}\},$ where
$\nu_{1}\in(-\infty,0)$, $\nu_{2}\in(4\lambda,\infty)$,

e2) for $|\delta|>1$ $s_{d}(H)=\{\nu_{1},\nu_{2},\nu_{3}\},$ where
$\nu_{1} \! \in \! (-\infty,0)$, $\nu_{2} \! \in\! (4\lambda,\infty)$, $\nu_{3}=\kappa;$

f) if $\left(\alpha,\delta,\sigma\right)\in U(+,-),$ then

f1) for $|\delta|\leqslant1$ $s_{d}(H)=\{\nu_{1}\},$ where $\nu_{1}\in(-\infty,0)$,
and

f2) for $|\delta|>1$ $s_{d}(H)=\{\nu_{1},\nu_{2}\},$ where $\nu_{1}\in(-\infty,0),$$\nu_{2}=\kappa$;

g) geometric multiplicity of all eigenvalues is $1;$ the eigenvector
$f=\{f_{n}\}\in l_{2}(\Z)$ corresponding to the eigenvalue $\kappa$
is an odd function ($f_{n}=-f_{-n}$) and the eigenvectors $f=\{f_{n}\}\in l_{2}(\Z)$
corresponding to the eigenvalues $\nu_{1},\nu_{2}$ are even functions
($f_{n}=f_{-n}$).
\end{theorem}

For $\alpha=\lambda_{1} / \lambda =-1$ one can find eigenvalues
in an explicit form.

\begin{theorem}\label{th2}Let $\lambda>0$ and $\kappa$ be defined
by (\ref{kappa}). For $\alpha=\lambda_{1} / \lambda =-1$ 
\begin{itemize}
\item if $|\delta|>1,$ $|\sigma|>1,$ $\mu\neq\kappa,$ then $s_{d}(H)=\{\mu,\kappa\}$ 
\item if $|\delta|\leq1,$ $|\sigma|>1,$ then $s_{d}(H)=\{\mu\}$ 
\item if $|\delta|>1,$ $|\sigma|\leq1,$ then $s_{d}(H)=\{\kappa\}$ 
\item if $|\delta|\leq1,$ $|\sigma|\leq1,$ then $s_{d}(H)=\{\emptyset\}$ 
\item the eigenvalue $\kappa$ is of the geometrical multiplicity $2$ and
has two linearly independent eigenvectors one of which is an odd function
and the other is an even function. 
\item if $\mu=\kappa,$ $|\delta|>1,$ $|\sigma|>1,$ there is only one
eigenvalue of the geometrical multiplicity $2$ with the same eigenvectors
as in the previous item. 
\end{itemize}
\end{theorem}

Note that for $\alpha=\lambda_{1} / \lambda =-1$ there exists
the eigenvalue $\mu$ with the eigenvector $e_{0}.$ By definition
(\ref{H5}), one can obtain
\begin{align*}
He_{0} & =H_{0}e_{0}+H_{1}e_{0}=\\
 & =-\lambda(e_{1}-2e_{0}+e_{-1})-\lambda_{1}(e_{1}-2e_{0}+e_{-1})+\mu e_{0}=\mu e_{0}
\end{align*}
So, if $\mu\in[0,4\lambda]$, then there exists the eigenvalue $\nu=\mu$
belonging to the essential spectrum of Hamiltonian $H.$

\section{Proof of main results}

For $f\in l_{2}(\Z)$ we have 
\begin{align}
Hf & =\sum_{n\in Z}\lambda\left(-f_{n+1}+2f_{n}-f_{n-1}\right)e_{n}+(\mu_{1}f_{-1}-\lambda_{1}f_{0})e_{-1}+\nonumber \\
 & +\left((2\lambda_{1}+\mu)f_{0}-\lambda_{1}f_{1}-\lambda_{1}f_{-1})\right)e_{0}+(\mu_{1}f_{1}-\lambda_{1}f_{0})e_{1}\label{T1}
\end{align}

Let us consider the Hilbert space $L_{2}(\S)$ of square integrable
functions defined on the unit circle $\S$. The elements of $L_{2}(\S)$
we will denote by $\hat{f}.$ It is known that any two separable Hilbert
spaces are isomorphic, so we have $l_{2}(\Z)\cong L_{2}(\S)$.

Consider an isomorphism $U\,:\,l_{2}(\Z)\rightarrow L_{2}(\S)$ such
that $e_{n}\longleftrightarrow\frac{1}{\sqrt{2\pi}}e^{i\varphi n}.$
Then $U$ is an unitary operator and for $f=\sum_{n\in Z}f_{n}e_{n}$
we have
\[
Uf=\sum_{n\in Z}f_{n}\frac{1}{\sqrt{2\pi}}e^{i\varphi n}
\]
Introduce operator $\hat{H}=UHU^{-1}\,:\,L_{2}(\S)\rightarrow L_{2}(\S)$
which is unitarily equivalent to $H.$ Using (\ref{T1}) we get
\begin{multline}
\hat{H}\hat{f}=\sum_{n\in Z}\lambda\left(-f_{n+1}+2f_{n}-f_{n-1}\right)\frac{1}{\sqrt{2\pi}}e^{i\varphi n}+(\mu_{1}f_{-1}-\lambda_{1}f_{0})\frac{1}{\sqrt{2\pi}}e^{-i\varphi}+\\
+\left((2\lambda_{1}+\mu)f_{0}-\lambda_{1}f_{1}-\lambda_{1}f_{-1})\right)\frac{1}{\sqrt{2\pi}}+(\mu_{1}f_{1}-\lambda_{1}f_{0})\frac{1}{\sqrt{2\pi}}e^{i\varphi}=\\
=\sum_{n\in Z}\lambda\left(-f_{n+1}\frac{1}{\sqrt{2\pi}}e^{i\varphi(n+1)}e^{-i\varphi}+2f_{n}\frac{1}{\sqrt{2\pi}}e^{i\varphi n}-f_{n-1}\frac{1}{\sqrt{2\pi}}e^{i\varphi(n-1)}e^{i\varphi}\right)-\\
-\frac{\lambda_{1}}{\sqrt{2\pi}}f_{0}(e^{-i\varphi}+e^{i\varphi}-2)-\frac{\lambda_{1}}{\sqrt{2\pi}}(f_{-1}+f_{1})+\frac{\mu}{\sqrt{2\pi}}f_{0}+\frac{\mu_{1}}{\sqrt{2\pi}}(f_{-1}e^{-i\varphi}+f_{1}e^{i\varphi})=\\
=-\lambda\left(e^{-i\varphi}+e^{i\varphi}-2\right)\hat{f}-\frac{\lambda_{1}}{\sqrt{2\pi}}\left(f_{-1}+f_{1}+f_{0}(e^{-i\varphi}+e^{i\varphi}-2)\right)+\\
+\frac{\mu}{\sqrt{2\pi}}f_{0}+\frac{\mu_{1}}{\sqrt{2\pi}}(f_{-1}e^{-i\varphi}+f_{1}e^{i\varphi})
\end{multline}
Thus, 
\begin{multline}
\hat{H}\hat{f}=-2\lambda\left(\cos\varphi-1\right)\hat{f}-\frac{\lambda_{1}}{\sqrt{2\pi}}\left(f_{-1}+f_{1}+2f_{0}(\cos\varphi-1)\right)+\\
+\frac{\mu}{\sqrt{2\pi}}f_{0}+\frac{\mu_{1}}{\sqrt{2\pi}}(f_{-1}e^{-i\varphi}+f_{1}e^{i\varphi})\label{T2}
\end{multline}

Because of unitarily equivalence of operators $\hat{H}$ and $H$,
their point spectra are identical.

If $\hat{f}$ is an eigenfunction with eigenvalue $\nu,$ i.e. $\hat{H}\hat{f}=\nu\hat{f,}$
then by (\ref{T2}) we have 
\begin{multline*}
\nu\hat{f}=-2\lambda\left(\cos\varphi-1\right)\hat{f}-\frac{\lambda_{1}}{\sqrt{2\pi}}\left(f_{-1}+f_{1}+2f_{0}(\cos\varphi-1)\right)+\\
+\frac{\mu}{\sqrt{2\pi}}f_{0}+\frac{\mu_{1}}{\sqrt{2\pi}}(f_{-1}e^{-i\varphi}+f_{1}e^{i\varphi})
\end{multline*}
and, hence,
\begin{multline*}
\hat{f}=\frac{1}{\sqrt{2\pi}}\frac{-\frac{\lambda_{1}}{2\lambda}\left(f_{-1}+f_{1}+2f_{0}(\cos\varphi-1)\right)+\frac{\mu}{2\lambda}f_{0}+\frac{\mu_{1}}{2\lambda}(f_{-1}e^{-i\varphi}+f_{1}e^{i\varphi})}{\cos\varphi+\frac{\nu}{2\lambda}-1}=\\[4pt]
=\frac{1}{\sqrt{2\pi}}\frac{-\frac{\alpha}{2}\left(f_{-1}+f_{1}\right)+(\sigma-\alpha\cos\varphi)f_{0}+\frac{\delta}{2}(f_{-1}e^{-i\varphi}+f_{1}e^{i\varphi})}{\cos\varphi+\frac{\nu}{2\lambda}-1}
\end{multline*}
Since 
\[
f_{-1}e^{-i\varphi}+f_{1}e^{i\varphi}=\left(f_{-1}+f_{1}\right)\cos\varphi+i\left(-f_{-1}+f_{1}\right)\sin\varphi
\]
we get 
\begin{multline*}
\hat{f}=\frac{1}{\sqrt{2\pi}}\frac{-\frac{\alpha}{2}\left(f_{-1}+f_{1}\right)+(\sigma-\alpha\cos\varphi)f_{0}}{\cos\varphi+\frac{\nu}{2\lambda}-1}+\\[4pt]
+\frac{1}{\sqrt{2\pi}}\frac{\frac{\delta}{2}\left(\left(f_{-1}+f_{1}\right)\cos\varphi+i\left(-f_{-1}+f_{1}\right)\sin\varphi\right)}{\cos\varphi+\frac{\nu}{2\lambda}-1}=\\[4pt]
=\frac{1}{\sqrt{2\pi}}\frac{\frac{f_{-1}+f_{1}}{2}\left(-\alpha+\delta\cos\varphi\right)+(\sigma-\alpha\cos\varphi)f_{0}+i\delta\frac{f_{1}-f_{-1}}{2}\sin\varphi}{\cos\varphi+\frac{\nu}{2\lambda}-1}
\end{multline*}
Put
\begin{equation}
\gamma=\frac{\nu}{2\lambda}-1\label{T6}
\end{equation}
It is evident that $\gamma\in[-1,1]\Longleftrightarrow\nu\in[0,4\lambda].$
Then 
\begin{equation}
\hat{f}=\frac{1}{\sqrt{2\pi}}\frac{\frac{f_{-1}+f_{1}}{2}\left(\delta\cos\varphi-\alpha\right)+(\sigma-\alpha\cos\varphi)f_{0}+i\delta\frac{f_{1}-f_{-1}}{2}\sin\varphi}{\cos\varphi+\gamma}\label{eq:19}
\end{equation}
If $\gamma\notin[-1,1],$ then $\hat{f}\in L_{2}(\S).$

Applying the inverse Fourier transform we get 
\begin{align}
f_{k}&=\frac{1}{\sqrt{2\pi}}\int_{-\pi}^{\pi}\hat{f}\,e^{-ik\varphi}d\varphi= \label{four_tr}\\
&=\frac{1}{2\pi}\int_{-\pi}^{\pi}\frac{\frac{f_{-1}+f_{1}}{2}\left(\delta\cos\varphi-\alpha\right)+(\sigma-\alpha\cos\varphi)f_{0}+i\delta\frac{f_{1}-f_{-1}}{2}\sin\varphi}{\cos\varphi+\gamma}e^{-ik\varphi}d\varphi \nonumber 
\end{align}
for all $k\in\Z$. It follows, that the coordinates $f_{0},f_{1},f_{-1}$
of eigenvector $f=\{f_{k}\}$ satisfy the following homogeneous linear
system:
\begin{equation}
f_{0}=\frac{1}{2\pi}\int_{-\pi}^{\pi}\frac{\frac{f_{-1}+f_{1}}{2}\left(\delta\cos\varphi-\alpha\right)+(\sigma-\alpha\cos\varphi)f_{0}+i\delta\frac{f_{1}-f_{-1}}{2}\sin\varphi}{\cos\varphi+\gamma}d\varphi\label{T3}
\end{equation}
\begin{equation}
f_{1}=\frac{1}{2\pi}\int_{-\pi}^{\pi}\frac{\frac{f_{-1}+f_{1}}{2}\left(\delta\cos\varphi-\alpha\right)+(\sigma-\alpha\cos\varphi)f_{0}+i\delta\frac{f_{1}-f_{-1}}{2}\sin\varphi}{\cos\varphi+\gamma}e^{-i\varphi}d\varphi\label{T4}
\end{equation}
\begin{equation}
f_{-1}=\frac{1}{2\pi}\int_{-\pi}^{\pi}\frac{\frac{f_{-1}+f_{1}}{2}\left(\delta\cos\varphi-\alpha\right)+(\sigma-\alpha\cos\varphi)f_{0}+i\delta\frac{f_{1}-f_{-1}}{2}\sin\varphi}{\cos\varphi+\gamma}e^{i\varphi}d\varphi\label{T4-1}
\end{equation}

Thus, we proved the following lemma:

\begin{lemma}\label{l1}
Real $\nu\notin[0,4\lambda]$ is an eigenvalue of the Hamiltonian
$H$ iff linear system (\ref{T3})--(\ref{T4-1}) has a nontrivial
solution for $\gamma=\frac{\nu}{2\lambda}-1.$ The geometric multiplicity
of eigenvalue $\nu$ is equal to $3-r,$ where $r$ is the rank of
linear system (\ref{T3})--(\ref{T4-1}).
\end{lemma}

Now we simplify integrals in equations (\ref{T3})--(\ref{T4-1}).
Let us denote 
\[
f_{1}^{\prime}=\frac{f_{-1}+f_{1}}{2},\quad f_{-1}^{\prime}=\frac{f_{1}-f_{-1}}{2}
\]
We have from equation (\ref{T3})
\begin{multline*}
f_{0}=\int_{-\pi}^{\pi}\frac{f_{1}^{\prime}\left(\delta\cos\varphi-\alpha\right)+(\sigma-\alpha\cos\varphi)f_{0}+i\delta f_{-1}^{\prime}\sin\varphi}{\cos\varphi+\gamma}d\varphi=\\
=\int_{-\pi}^{\pi}\frac{f_{1}^{\prime}\left(\delta\cos\varphi-\alpha\right)+(\sigma-\alpha\cos\varphi)f_{0}}{\cos\varphi+\gamma}d\varphi
\end{multline*}
since 
\[
\int_{-\pi}^{\pi}\frac{i\frac{\delta}{2}\left(-f_{-1}+f_{1}\right)\sin\varphi}{\cos\varphi+\gamma}d\varphi=0
\]
because of the integrand is an odd function.

Further, from equation (\ref{T4}) we get 
\begin{multline*}
f_{1}=\int_{-\pi}^{\pi}\frac{f_{1}^{\prime}\left(\delta\cos\varphi-\alpha\right)+(\sigma-\alpha\cos\varphi)f_{0}}{\cos\varphi+\gamma}\left(\cos\varphi-i\sin\varphi\right)d\varphi+\\
+\int_{-\pi}^{\pi}\frac{i\delta f_{-1}^{\prime}\sin\varphi}{\cos\varphi+\gamma}\left(\cos\varphi-i\sin\varphi\right)d\varphi
\end{multline*}
Because of the integrands are odd functions we have 
\begin{align*}
\int_{-\pi}^{\pi}\frac{f_{1}^{\prime}\left(\delta\cos\varphi-\alpha\right)+(\sigma-\alpha\cos\varphi)f_{0}}{\cos\varphi+\gamma}i\sin\varphi d\varphi & =0\\
\int_{-\pi}^{\pi}\frac{\sin\varphi}{\cos\varphi+\gamma}\cos\varphi d\varphi & =0
\end{align*}

So 
\begin{align*}
f_{1}=\int_{-\pi}^{\pi}\frac{f_{1}^{\prime}\left(\delta\cos\varphi-\alpha\right)+(\sigma-\alpha\cos\varphi)f_{0}}{\cos\varphi+\gamma}\cos\varphi d\varphi+
    \int_{-\pi}^{\pi}\frac{\delta f_{-1}^{\prime}\sin^{2}\varphi}{\cos\varphi+\gamma}d\varphi
\end{align*}
Similarly, one can show that 
\begin{align*}
f_{-1}=\int_{-\pi}^{\pi}\frac{f_{1}^{\prime}\left(\delta\cos\varphi-\alpha\right)+(\sigma-\alpha\cos\varphi)f_{0}}{\cos\varphi+\gamma}\cos\varphi d\varphi
    -\int_{-\pi}^{\pi}\frac{\delta f_{-1}^{\prime}\sin^{2}\varphi}{\cos\varphi+\gamma}d\varphi
\end{align*}
Hence, system (\ref{T3})--(\ref{T4-1}) can be rewritten as follows
\begin{equation}
f_{0}=\frac{1}{2\pi}\int_{-\pi}^{\pi}\frac{f_{1}^{\prime}\left(\delta\cos\varphi-\alpha\right)+(\sigma-\alpha\cos\varphi)f_{0}}{\cos\varphi+\gamma}d\varphi\label{eq:T1}
\end{equation}
\begin{align}
f_{1} =\frac{1}{2\pi}\int_{-\pi}^{\pi}\frac{f_{1}^{\prime}\left(\delta\cos\varphi-\alpha\right)+(\sigma-\alpha\cos\varphi)f_{0}}{\cos\varphi+\gamma}\cos\varphi d\varphi
  +\frac{1}{2\pi}\int_{-\pi}^{\pi}\frac{\delta f_{-1}^{\prime}\sin^{2}\varphi}{\cos\varphi+\gamma}d\varphi\label{eq:T2}
\end{align}
\begin{align}
f_{-1} &=\frac{1}{2\pi}\int_{-\pi}^{\pi}\frac{f_{1}^{\prime}\left(\delta\cos\varphi-\alpha\right)+(\sigma-\alpha\cos\varphi)f_{0}}{\cos\varphi+\gamma}\cos\varphi d\varphi
-\frac{1}{2\pi}\int_{-\pi}^{\pi}\frac{\delta f_{-1}^{\prime}\sin^{2}\varphi}{\cos\varphi+\gamma}d\varphi\label{eq:T3}
\end{align}
Adding equations (\ref{eq:T2}) and (\ref{eq:T3}), subtracting equation
(\ref{eq:T3}) from (\ref{eq:T2}) and dividing the resulting equations
by $2$ we come to the linear system with respect to variables $f_{0},f_{1}^{\prime},f_{-1}^{\prime}$:
\[
f_{0}=\frac{1}{2\pi}\int_{-\pi}^{\pi}\frac{(\sigma-\alpha\cos\varphi)f_{0}+\left(\delta\cos\varphi-\alpha\right)f_{1}^{\prime}}{\cos\varphi+\gamma}d\varphi
\]

\[
f_{1}^{\prime}=\frac{1}{2\pi}\int_{-\pi}^{\pi}\frac{\left(\sigma-\alpha\cos\varphi\right)\cos\varphi f_{0}+\left(\delta\text{cos}\varphi-\alpha\right)\cos\varphi f_{1}^{\prime}}{\cos\varphi+\gamma}d\varphi
\]

\[
f_{-1}^{\prime}=f_{-1}^{\prime}\frac{1}{2\pi}\int_{-\pi}^{\pi}\frac{\delta\sin^{2}\varphi}{\cos\varphi+\gamma}d\varphi
\]

Finely, we get the system which is equivalent to the original system
(\ref{T3})--(\ref{T4-1}):
\begin{equation}
f_{0}\Biggl(1-\frac{1}{2\pi}\int_{-\pi}^{\pi}\frac{\sigma-\alpha\cos\varphi}{\cos\varphi+\gamma}d\varphi\Biggr)+\frac{f_{1}^{\prime}}{2\pi}\int_{-\pi}^{\pi}\frac{\alpha-\delta\text{cos}\varphi}{\cos\varphi+\gamma}d\varphi=0\label{T31}
\end{equation}

\begin{equation}
  -\frac{f_{0}}{2\pi}\int_{-\pi}^{\pi}\frac{\left(\sigma-\alpha\cos\varphi\right)\cos\varphi}{\cos\varphi+\gamma}d\varphi+f_{1}^{\prime}
  \Biggl(1+\frac{1}{2\pi}\int_{-\pi}^{\pi}\frac{\left(\alpha-\delta\text{cos}\varphi\right)\text{cos}\varphi}{\cos\varphi+\gamma}d\varphi\Biggr)=0  \label{T41}
\end{equation}

\begin{equation}
f_{-1}^{\prime}\Biggl(\frac{1}{2\pi}\int_{-\pi}^{\pi}\frac{\delta\left(1-\cos^{2}\varphi\right)}{\cos\varphi+\gamma}d\varphi-1\Biggr)=0\label{T51}
\end{equation}
Define the following functions
\begin{eqnarray*}
g(x) & = & \frac{1}{2\pi}\int_{-\pi}^{\pi}\frac{d\varphi}{\cos\varphi+x}\\
h(x) & = & -\frac{1}{2\pi}\int_{-\pi}^{\pi}\frac{\cos\varphi d\varphi}{\cos\varphi+x}\\
v(x) & = & \frac{1}{2\pi}\int_{-\pi}^{\pi}\frac{\cos^{2}\varphi d\varphi}{\cos\varphi+x}
\end{eqnarray*}
where $|x|>1.$ By lemma \ref{Lemma2}
\begin{equation}
g(x)=\frac{1}{x\sqrt{1-x^{-2}}},\label{L1}
\end{equation}
\begin{equation}
h(x)=xg(x)-1,\:v(x)=xh(x),\label{L3}
\end{equation}
\begin{equation}
g(x)v(x)=h^{2}(x)+h(x)\label{L4}
\end{equation}
Let us express coefficients of linear system (\ref{T31})\textendash (\ref{T51})
in terms of functions $g,h,v:$
\begin{equation}
f_{0}(1-\alpha h(\gamma)-\sigma g(\gamma))+f_{1}^{\prime}(\alpha g(\gamma)+\delta h(\gamma))=0,\label{S1}
\end{equation}
\begin{equation}
f_{0}(\sigma h(\gamma)+\alpha v(\gamma))+f_{1}^{\prime}\left(1-\alpha h(\gamma)-\delta v(\gamma)\right)=0\label{S2}
\end{equation}
\begin{equation}
f_{-1}^{\prime}\left(\delta\left(g(\gamma)-v(\gamma)\right)-1\right)=0\label{S3}
\end{equation}
Coefficients of this system are defined only for $|\gamma|>1.$ We
need to find such $\gamma$ that system (\ref{S1})\textendash (\ref{S3})
has nontrivial solution. A homogeneous linear system has non trivial
solution iff its determinant equals zero. Denote by $D^{\prime}(\alpha,\delta,\sigma,\gamma)$
the determinant of this system and let $D(\alpha,\delta,\sigma,\gamma)$
be the determinant of system (\ref{S1})\textendash (\ref{S2}), consisting
of the first two equations. We have 
\begin{equation}
D^{\prime}(\alpha,\delta,\sigma,\gamma)=D(\alpha,\delta,\sigma,\gamma)\left(\delta\left(g(\gamma)-v(\gamma)\right)-1\right)\label{eq:deter}
\end{equation}
So in order to determine eigenvalues we need to find roots of the
equation $D^{\prime}(\alpha,\delta,\sigma,\gamma)=0.$ Simple algebra
gives
\begin{align*}
D(\alpha,\delta,\sigma,\gamma) & =(1-\alpha h(\gamma)-\sigma g(\gamma))\left(1-\alpha h(\gamma)-\delta v(\gamma)\right)-\\
 & -(\alpha g(\gamma)+\delta h(\gamma))(\sigma h(\gamma)+\alpha v(\gamma))=\\
 & =(1-\alpha h(\gamma))^{2}-\delta v(\gamma)+\alpha h(\gamma)\delta v(\gamma)-\\
 & -\sigma g(\gamma)\left(1-\delta v(\gamma)\right)+\sigma g(\gamma)\alpha h(\gamma)-\\
 & -\alpha g(\gamma)\sigma h(\gamma)-\alpha^{2}g(\gamma)v(\gamma)-\delta\sigma h^{2}(\gamma)-\delta h(\gamma)\alpha v(\gamma)=\\
 & =(1-\alpha h(\gamma))^{2}-\delta v(\gamma)-\sigma g(\gamma)\left(1-\delta v(\gamma)\right)-\\
 & -\alpha^{2}g(\gamma)v(\gamma)-\delta\sigma h^{2}(\gamma)=\\
 & =(1-\alpha h(\gamma))^{2}-\delta\sigma h^{2}(\gamma)-\sigma g(\gamma)-\left(\alpha^{2}-\sigma\delta\right)g(\gamma)v(\gamma)-\delta v(\gamma)
\end{align*}
By (\ref{L4}) we have $g(\gamma)v(\gamma)=h^{2}(\gamma)+h(\gamma)$
and, hence,
\begin{align*}
D(\alpha,\delta,\sigma,\gamma) & =(1-\alpha h(\gamma))^{2}-\delta\sigma h^{2}(\gamma)-\sigma g(\gamma)-\left(\alpha^{2}-\sigma\delta\right)h^{2}(\gamma)-\\
 & -\left(\alpha^{2}-\sigma\delta\right)h(\gamma)-\delta v(\gamma)\\
 & =1-2\alpha h(\gamma)+\alpha^{2}h^{2}(\gamma)-\delta\sigma h^{2}(\gamma)-\sigma g(\gamma)-\alpha^{2}h^{2}(\gamma)+\\
 & +\sigma\delta h^{2}(\gamma)-\alpha^{2}h(\gamma)+\sigma\delta h(\gamma)-\delta v(\gamma)=\\
 & =1-\sigma g(\gamma)-\left(\alpha^{2}+2\alpha-\sigma\delta\right)h(\gamma)-\delta v(\gamma)
\end{align*}

Remark that $h(\gamma)\neq0$ for $|\gamma|>1.$ Let us divide $D(\alpha,\delta,\sigma,\gamma)$
by $h(\gamma)$. Using (\ref{L3}), we obtain 
\[
D(\alpha,\delta,\sigma,\gamma)=h(\gamma)\Bigl(\frac{1}{h(\gamma)}-\sigma\frac{g(\gamma)}{h(\gamma)}-\alpha^{2}-2\alpha+\sigma\delta-\delta\gamma\Bigr)
\]
Put 
\begin{equation}
l(\gamma)=\gamma\left(1+\sqrt{1-\gamma^{-2}}\right)=\begin{cases}
\gamma+\sqrt{\gamma^{2}-1} & \gamma\geq1\\
\gamma-\sqrt{\gamma^{2}-1} & \gamma\leq-1
\end{cases}\label{lg}
\end{equation}
By (\ref{L1}), we have 
\begin{align*}
\frac{1}{h(\gamma)}&=\frac{\sqrt{1-\gamma^{-2}}}{1-\sqrt{1-\gamma^{-2}}}=\frac{\sqrt{1-\gamma^{-2}}-1+1}{1-\sqrt{1-\gamma^{-2}}}=\\[4pt]
&=-1+\frac{1}{1-\sqrt{1-\gamma^{-2}}}\\[4pt]
\frac{g(\gamma)}{h(\gamma)}&=\frac{1}{\gamma\bigl(1-\sqrt{1-\gamma^{-2}}\bigr)}
\end{align*}
Since 
\begin{align*}
\frac{1}{1-\sqrt{1-\gamma^{-2}}} & =\frac{1+\sqrt{1-\gamma^{-2}}}{\bigl(1-\sqrt{1-\gamma^{-2}}\bigr)\bigl(1+\sqrt{1-\gamma^{-2}}\bigr)}\\[3pt]
 & =\gamma^{2}\bigl(1+\sqrt{1-\gamma^{-2}}\bigr)=\gamma l(\gamma)
\end{align*}
we get 
\begin{align*}
\frac{1}{h(\gamma)}  =-1+\gamma l(\gamma), \qquad
\frac{g(\gamma)}{h(\gamma)}  =l(\gamma)
\end{align*}
and, hence, 
\begin{align}
D(\alpha,\delta,\sigma,\gamma)&=h(\gamma)\left(\gamma l(\gamma)-\sigma l(\gamma)-\delta\left(\gamma-\sigma\right)-\left(\alpha+1\right)^{2}\right)=\nonumber \\
&=h^{-1}(\gamma)\bigl(\left(\gamma-\sigma\right)\left(l(\gamma)-\delta\right)-\left(\alpha+1\right)^{2}\bigr)\label{eq:deter_0}
\end{align}

Consider the second factor in (\ref{eq:deter}). By (\ref{L3}) we
have 
\begin{align*}
g(\gamma)-v(\gamma) & =g(\gamma)-\gamma h(\gamma)=g(\gamma)-\gamma\left(\gamma g(\gamma)-1\right)\\
 & =-g(\gamma)\gamma^{2}\left(1-\gamma^{-2}\right)+\gamma
\end{align*}
Now using (\ref{L1}) we get 
\begin{multline}
g(\gamma)-v(\gamma)=-\frac{\gamma^{2}\left(1-\gamma^{-2}\right)}{\gamma\sqrt{1-\gamma^{-2}}}+\gamma=\gamma\left(1-\sqrt{1-\gamma^{-2}}\right)=\\
=\frac{\gamma\left(1-\sqrt{1-\gamma^{-2}}\right)\left(1+\sqrt{1-\gamma^{-2}}\right)}{1+\sqrt{1-\gamma^{-2}}}=\\
=\frac{1}{\gamma\left(1+\sqrt{1-\gamma^{-2}}\right)}=\frac{1}{l(\gamma)}\qquad\;\label{eq:g-v}
\end{multline}
and the second factor in (\ref{eq:deter}) is 
\[
\delta\left(g(\gamma)-v(\gamma)\right)-1=\frac{\delta}{l(\gamma)}-1=\frac{-l(\gamma)+\delta}{l(\gamma)}
\]
where $l(\gamma)$ is defined by (\ref{lg}). According to (\ref{eq:deter})
\begin{equation}
D^{\prime}(\alpha,\delta,\sigma,\gamma)=-h(\gamma)l^{-1}(\gamma)\left(\left(\gamma-\sigma\right)\left(l(\gamma)-\delta\right)-\left(\alpha+1\right)^{2}\right)\left(l(\gamma)-\delta\right)\label{eq:deter_1}
\end{equation}
Thus $D^{\prime}(\alpha,\delta,\sigma,\gamma)=0$ iff or $l(\gamma)-\delta=0$
or $\left(\gamma-\sigma\right)\left(l(\gamma)-\delta\right)-\left(\alpha+1\right)^{2}=0\text{.}$

Note that $|l(\gamma)|\geq1$ and the equation $l(\gamma)-\delta=0$
has a unique root $\gamma_{\delta}=\frac{\delta^{2}+1}{2\delta}$
iff $|\delta|\geq1$ ($|\gamma|=1$ iff $|\delta|=1$). Hence, in
case of $|\delta|>1$ we have eigenvalue 
\[
\kappa=2\lambda\left(\frac{\delta^{2}+1}{2\delta}+1\right)=\frac{\lambda(\delta+1)^{2}}{\delta}=\frac{(\mu_{1}+\lambda)^{2}}{\mu_{1}}
\]

Let us find an eigenvector corresponding to this eigenvalue. To satisfy
linear system (\ref{S1})--(\ref{S3}) one can put $f_{1}^{\prime}=f_{1}+f_{-1}=0$,
$f_{0}=0,$ $f_{-1}^{\prime}=f_{1}-f_{-1}=2f_{1}$ and for all $|k|>1$
by (\ref{four_tr}) we have
\begin{align}
f_{k} & =\frac{1}{2\pi}\int_{-\pi}^{\pi}\frac{\frac{\delta}{2}(f_{-1}e^{-i\varphi}+f_{1}e^{i\varphi})}{\cos\varphi+\gamma}e^{-ik\varphi}d\varphi\nonumber \\
 & =\frac{1}{2\pi}\int_{-\pi}^{\pi}\frac{\frac{\delta}{2}f_{1}(-e^{-i\varphi}+e^{i\varphi})}{\cos\varphi+\gamma}e^{-ik\varphi}d\varphi=\nonumber \\
 & =\frac{1}{2\pi}\int_{-\pi}^{\pi}\frac{\delta if_{1}\sin\varphi}{\cos\varphi+\gamma}\left(\cos k\varphi-i\sin k\varphi\right)d\varphi=\nonumber \\
 & =\frac{\delta f_{1}}{2\pi}\int_{-\pi}^{\pi}\frac{\sin\varphi\sin k\varphi}{\cos\varphi+\gamma}d\varphi=\nonumber \\
 & =\frac{\delta f_{1}}{4\pi}\int_{-\pi}^{\pi}\frac{\cos\left((k-1)\varphi\right)-\cos\left((k+1)\varphi\right)}{\cos\varphi+\gamma}d\varphi\label{odd_eig}
\end{align}

We see that $f_{k}$ is an odd function of $k:$ $f_{k}=-f_{-k}.$

Let $\alpha\neq-1.$ Define the function 
\begin{equation}
F(\gamma)=\left(\gamma-\sigma\right)\left(l(\gamma)-\delta\right)\label{F10-1}
\end{equation}
for $|\gamma|\geq1.$ As $\gamma\to\pm\infty$ $F(\gamma)\to\infty.$

Consider the equation 
\begin{equation}
F(\gamma)=\left(\alpha+1\right)^{2}\label{D2}
\end{equation}
One can see (since $l(\pm1)=\pm1$) 
\begin{align*}
c_{1}= & c_{1}(\alpha,\delta,\sigma)=(\delta-1)(\sigma-1)-(\alpha+1)^{2}=F(1)-(\alpha+1)^{2}\\
c_{2}= & c_{2}(\alpha,\delta,\sigma)=(\delta+1)(\sigma+1)-(\alpha+1)^{2}=F(-1)-(\alpha+1)^{2}
\end{align*}
The equation $l(\gamma)=\delta$ has a unique root 
\[
\gamma_{\delta}=\frac{\delta^{2}+1}{2\delta}=\frac{1}{2}\Bigl(\delta+\frac{1}{\delta}\Bigr)
\]
iff $|\delta|\geq1.$

\smallskip

a) Let $\left(\alpha,\delta,\sigma\right)\in U_{1}(+,+)=\left\{ \left(\alpha,\delta,\sigma\right)\in{\bf R}^{3}:c_{1}>0,c_{2}>0,\sigma>1,\delta>1\right\} $

The equation $F(\gamma)=0$ has two roots $\sigma$ and $\gamma_{\delta}$
which are greater than $1.$ Put 
\[
\gamma_{min}=\min(\sigma,\gamma_{\delta}),\gamma_{max}=\max(\sigma,\gamma_{\delta})
\]
Function $F(\gamma)$ is strictly decreasing and positive on the interval
$(1,\gamma_{min})$ and takes all values in the interval $(0,F(1));$
$F(\gamma)$ is strictly increasing and positive on the interval $(\gamma_{max},\infty)$
and takes all positive real values. For $\gamma\in(\gamma_{min},\gamma_{max})$
we have $F(\gamma)<0.$ The condition $c_{1}>0$ is equivalent to
$F(1)>(\alpha+1)^{2}.$ So equation $F(\gamma)=(\alpha+1)^{2}$ has
exactly two solutions $\gamma_{1}\in(1,\gamma_{min}),\gamma_{2}\in(\gamma_{max},\infty).$

For $\gamma<-1$ there is no solution as $F(-1)>F(1)>(\alpha+1)^{2}$
and $F(\gamma)$ is strictly decreasing on the interval $(-\infty,-1).$

\smallskip

b) Let 
$$
\left(\alpha,\delta,\sigma\right)\in U_{0}(+,+)=\left\{ \left(\alpha,\delta,\sigma\right)\in{\bf R}^{3}:c_{1}\geqslant0,c_{2}\geqslant0,|\sigma|\leqslant1,|\delta|\leqslant1\right\}
$$
In this case the function $F(\gamma)$ is strictly decreasing on the
interval $(-\infty,-1)$ and takes all values in the interval $(F(-1),\infty)$;
for $\gamma\in(1,\infty)$ $F(\gamma)$ is strictly increasing and
takes all values in the interval $(F(1),\infty)$;

It follows from $c_{1}\geqslant0,c_{2}\geqslant0$ that $F(-1)\geq(\alpha+1)^{2}$
and $F(1)\geq(\alpha+1)^{2}.$ So the equation $F(\gamma)=(\alpha+1)^{2}$
has no solutions satisfying condition $|\gamma|>1.$

\smallskip

The proof of c) is similar to the proof of a).

\smallskip

d) Let $\left(\alpha,\delta,\sigma\right)\in U(-,+)=\left\{ \left(\alpha,\delta,\sigma\right)\in{\bf R}^{3}:c_{1}<0,c_{2}\geqslant0\right\} \cup$

$\cup\left\{ \left(\alpha,\delta,\sigma\right)\in{\bf R}^{3}:c_{1}=0,c_{2}>0,\sigma>1,\delta>1\right\} .$
If $c_{1}<0,c_{2}\geqslant0$ then we have
\[
F(-1)\geqslant(\alpha+1)^{2}>F(1)
\]

The condition $c_{1}<0$ implies that equation $F(\gamma)=\left(\alpha+1\right)^{2}$
has a unique solution satisfying condition $\gamma>1.$ Indeed, two
cases are possible. In the first one for $\gamma\in(1,\infty)$ $F(\gamma)$
is strictly increasing taking all values in the interval $(F(1),\infty)$
and, since $(\alpha+1)^{2}>F(1)$ the equation $(\ref{D2})$ has a
unique solution $\gamma>1.$ The second case is similar to item a)
except for now the condition $(\alpha+1)^{2}>F(1)$ holds and, hence,
the equation $(\ref{D2})$ has no solution belonging to the interval
$(1,\gamma_{\min}),$ but exactly one solution $\gamma>\gamma_{\max}.$

For $\gamma<-1$ there is no solution since $F(-1)\geqslant(\alpha+1)^{2}$
and $F(\gamma)$ is strictly decreasing for $\gamma<-1.$

So there exists a unique solution $\gamma\in(1,+\infty)$ satisfying
(\ref{D2}).

The case $\left\{ \left(\alpha,\delta,\sigma\right)\in{\bf R}^{3}:c_{1}=0,c_{2}>0,\sigma>1,\delta>1\right\} $
is similar to item a) except for now the condition $(\alpha+1)^{2}=F(1)$
holds and, hence, the equation $(\ref{D2})$ has a unique solution
$\gamma$ satisfying condition $\gamma>1$.

\smallskip

e) Let $\left(\alpha,\delta,\sigma\right)\in U(-,-)=\left\{ \left(\alpha,\delta,\sigma\right)\in{\bf R}^{3}:c_{1}<0,c_{2}<0\right\} $.
As well as for the item d) one can show that conditions $c_{1}<0,c_{2}<0$
imply the existence of exactly two solutions of equation $F(\gamma)=\left(\alpha+1\right)^{2}$
one of which is strictly greater than $1$ and the other is strictly
less than $-1.$

\smallskip

The proof of the item f) is similar to the proof of the item e).

\smallskip

To prove the item g) note, that the eigenvector corresponding to $\kappa$,
defined by (\ref{odd_eig}), is an odd function.

Let us find an eigenvector corresponding to the eigenvalue $\nu$,
where $\nu=2\lambda\left(\gamma+1\right)$ and $\gamma$ is a root
of equation (\ref{eq:deter_0}). Note, that if $\alpha\neq-1$ then
$l(\gamma)\neq\delta.$

It follows from (\ref{S3}) and (\ref{eq:g-v}), that $f_{-1}^{\prime}=\frac{-f_{-1}+f_{1}}{2}=0$
and, hence, $\frac{f_{-1}+f_{1}}{2}=f_{1}.$ By (\ref{four_tr}) we
have 
\begin{multline*}
f_{k}=\frac{1}{2\pi}\int_{-\pi}^{\pi}\frac{f_{1}\left(-\alpha+\delta\cos\varphi\right)+(\sigma-\alpha\cos\varphi)f_{0}}{\cos\varphi+\gamma}e^{-ik\varphi}d\varphi=\\
=\frac{1}{2\pi}\int_{-\pi}^{\pi}\frac{f_{1}\left(-\alpha+\delta\cos\varphi\right)+(\sigma-\alpha\cos\varphi)f_{0}}{\cos\varphi+\gamma}\left(\cos k\varphi-i\sin k\varphi\right)d\varphi
\end{multline*}
Because of the integrand is an odd function we have
\[
\frac{1}{2\pi}\int_{-\pi}^{\pi}\frac{f_{1}\left(-\alpha+\delta\cos\varphi\right)+(\sigma-\alpha\cos\varphi)f_{0}}{\cos\varphi+\gamma}i\sin k\varphi d\varphi=0
\]
and, hence 
\begin{equation}
f_{k}=\frac{1}{2\pi}\int_{-\pi}^{\pi}\frac{f_{1}\left(-\alpha+\delta\cos\varphi\right)+(\sigma-\alpha\cos\varphi)f_{0}}{\cos\varphi+\gamma}\cos k\varphi d\varphi\label{even-eig}
\end{equation}
It follows that $f_{k}$ is an even function of $k:$ $f_{k}=f_{-k}.$

This completes the proof of Theorem \ref{th1}.

\bigskip

In order to prove theorem \ref{th2} consider the case of $\alpha=-1.$
According to $(\ref{eq:deter_1})$ we have 
\[
D^{\prime}(\alpha,\delta,\sigma,\gamma)=-h(\gamma)l^{-1}(\gamma)\left(\gamma-\sigma\right)\left(l(\gamma)-\delta\right)^{2}
\]
and the equation for eigenvalues will take the simple form 
\[
\left(\gamma-\sigma\right)\left(l(\gamma)-\delta\right)^{2}=0
\]
We are looking for roots satisfying the condition $|\gamma|>1.$ So 
\begin{itemize}
\item if $|\delta|>1,$ $|\sigma|>1,$ then there are two roots $\gamma=\sigma$
and $\gamma=\frac{\delta^{2}+1}{2\delta}$ 
\item if $|\delta|\leq1,$ $|\sigma|>1,$ then there is one root $\gamma=\sigma$ 
\item if $|\delta|>1,$ $|\sigma|\leq1,$ then there is one root $\gamma=\frac{\delta^{2}+1}{2\delta}$ 
\item if $|\delta|\leq1,$ $|\sigma|\leq1,$ then there are no roots 
\end{itemize}
Remind that $\gamma=\frac{\nu}{2\lambda}-1$ and $\sigma=\frac{2\lambda_{1}+\mu}{2\lambda}.$
If $\alpha=\frac{\lambda_{1}}{\lambda}=-1$ then $\sigma=\frac{\mu}{2\lambda}-1$
and equality $\gamma=\sigma$ is equivalent to $\nu=\mu.$

According to (\ref{kappa}) equality $\gamma=\frac{\delta^{2}+1}{2\delta}$
is equivalent to $\nu=\kappa.$ By lemma \ref{l1} this eigenvalue
is of multiplicity $2$ since the rank of the system (\ref{T31})--(\ref{T51}) is equal to $1.$

If $\mu=\kappa,$ then there is only one eigenvalue of multiplicity
$2$ because of the rank of the system (\ref{T31})--(\ref{T51})
equals $1.$

The theorem is proved.

\section*{Appendix}

\def\thesection{A}

\subsection{Partition of parameter space}

Here we give an illustration of partition (\ref{U2}). We consider
sections of the parameter space ${\bf R}^{3}$ by planes $\alpha=c$
for various constants $c.$

\bigskip

\begin{figure}[h]
\centering
\includegraphics[width=9cm]{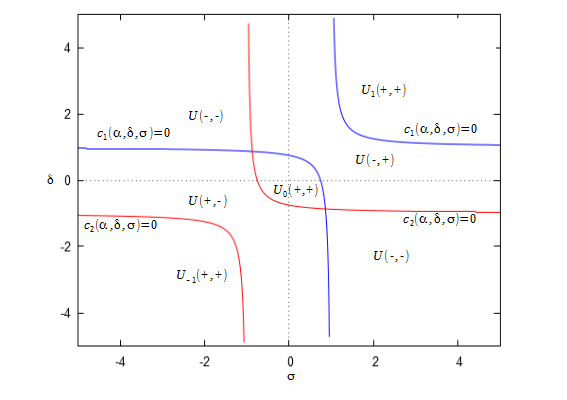}
\caption{Case of $|c+1|<1$}
\label{fig1}
\end{figure}
\begin{figure}[h]
\centering
\includegraphics[width=7cm]{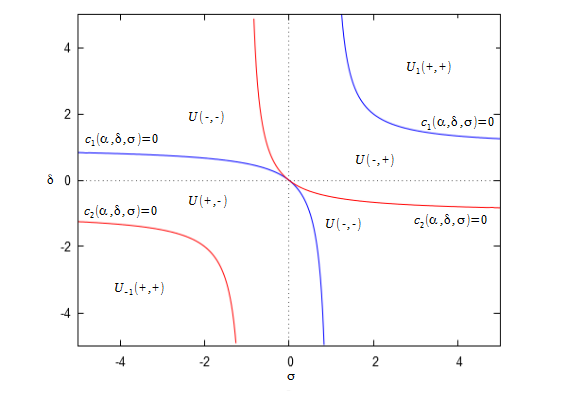}
\caption{Case of $|c+1|=1$}
\label{fig2}
\end{figure}
\begin{figure}[h]
\centering
\includegraphics[width=7cm]{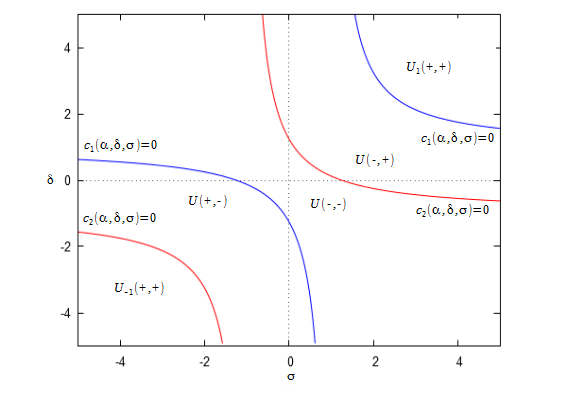}
\caption{Case of $|c+1|>1$}
\label{fig3}
\end{figure}

\subsection{Lemma \ref{Lemma2}}

Consider the functions
\begin{align}
g(x) & =\frac{1}{2\pi}\int_{-\pi}^{\pi}\frac{d\varphi}{\cos\varphi+x}\text{,}\label{L4-1}\\
h(x) & =-\frac{1}{2\pi}\int_{-\pi}^{\pi}\frac{\cos\varphi d\varphi}{\cos\varphi+x}\label{L4-2}\\
v(x) & =\frac{1}{2\pi}\int_{-\pi}^{\pi}\frac{\cos{}^{2}\varphi d\varphi}{\cos\varphi+x}\label{L4-3}
\end{align}
where $|x|>1.$

\begin{lemma}\label{Lemma2}For $|x|>1$
\begin{equation}
g(x)=\frac{1}{x\sqrt{1-x^{-2}}},\label{L5}
\end{equation}
\begin{equation}
h(x)=xg(x)-1,v(x)=xh(x),\label{L7}
\end{equation}
\begin{equation}
g(x)v(x)=h^{2}(x)+h(x)\label{L8}
\end{equation}
\end{lemma}
{\bf Proof}
Let us make change of variable $z=e^{i\varphi}$ in integral (\ref{L4-1}).
We get 
\begin{multline*}
g(x)=\frac{1}{2\pi}\int_{-\pi}^{\pi}\frac{d\varphi}{\cos\varphi+x}=\frac{1}{2\pi i}\int_{\Gamma}\frac{dz}{\left(\frac{z+z^{-1}}{2}+x\right)z}=\\
=\frac{1}{2\pi i}\int_{\Gamma}\frac{2dz}{z^{2}+2xz+1}=\frac{1}{2\pi i}\int_{\Gamma}\frac{2dz}{(z-a_{1})(z-a_{2})}
\end{multline*}
where $\Gamma=\{z:\:|z|=1\}$ is the unit circle and $a_{1}=-\left(x-\sqrt{x^{2}-1}\right)$,
$a_{2}=-\left(x+\sqrt{x^{2}-1}\right)$ are roots of square equation
$z^{2}+2xz+1=0.$ For $|x|>1$ only one of these roots is strictly
less than $1$ by module. If $x\text{>1}$, then $|a_{1}|=|x-\sqrt{x^{2}-1}|<1$
and 
\begin{multline*}
\frac{1}{2\pi i}\int_{\Gamma}\frac{2dz}{(z-a_{1})(z-a_{2})}=\\
={\bf Res}_{a_{1}}\frac{2}{(z-a_{1})(z-a_{2})}=\frac{1}{\sqrt{x^{2}-1}}=\frac{1}{x\sqrt{1-x^{-2}}}
\end{multline*}
If $x\text{<-1}$, then $|a_{2}|=|x+\sqrt{x^{2}-1}|<1$ and 
\begin{multline*}
\frac{1}{2\pi i}\int_{\Gamma}\frac{2dz}{(z-a_{1})(z-a_{2})}=\\
={\bf Res}_{a_{2}}\frac{2}{(z-a_{1})(z-a_{2})}=\frac{1}{-\sqrt{x^{2}-1}}=\frac{1}{x\sqrt{1-x^{-2}}}
\end{multline*}
For integral (\ref{L4-2}) we have
\[
h(x)=-\frac{1}{2\pi}\int_{-\pi}^{\pi}\frac{\cos\varphi d\varphi}{\cos\varphi+x}=\frac{x}{2\pi}\int_{-\pi}^{\pi}\frac{d\varphi}{\cos\varphi+x}-\frac{1}{2\pi}\int_{-\pi}^{\pi}d\varphi=xg(x)-1.
\]
Further, we get
\begin{align*}
v(x)&=\frac{1}{2\pi}\int_{-\pi}^{\pi}\frac{\cos^{2}\varphi+2x\cos\varphi+x^{2}-\left(2x\cos\varphi+x^{2}\right)}{\cos\varphi+x}d\varphi=\\
&=\frac{1}{2\pi}\int_{-\pi}^{\pi}\frac{\cos\varphi\left(\cos\varphi+x\right)-x\cos\varphi}{\cos\varphi+x}d\varphi=\\
&=\frac{1}{2\pi}\left(\int_{-\pi}^{\pi}\cos\varphi d\varphi-x\int_{-\pi}^{\pi}\frac{\cos\varphi d\varphi}{\cos\varphi+x}\right)=\\
&=-\frac{x}{2\pi}\int_{-\pi}^{\pi}\frac{\cos\varphi d\varphi}{\cos\varphi+x}=xh(x)
\end{align*}

Finally, 
\[
g(x)v(x)=g(x)xh(x)=\left(h(x)+1\right)h(x)=h^{2}(x)+h(x)
\]
This completes the proof of the lemma. 

\end{document}